\documentstyle[prd,aps,preprint]{revtex}

\begin{document}
\thispagestyle{empty}
\draft
\preprint{\font\fortssbx=cmssbx10 scaled \magstep2
\hbox to \hsize{
\hbox{\fortssbx University of Wisconsin - Madison}
\hfill$\vcenter{\hbox{\bf MADPH-96-932}
                \hbox{February 1996}}$ }}

\title{\vspace{.25in}
Composition of Primary Cosmic Rays Beyond the ``Knee'' \\
       from Emulsion Chamber Observations}
\author{C. G. S. Costa and F. Halzen}
\address{Department of Physics, University of Wisconsin,
         Madison, WI 53706, USA}
\author{J. Bellandi}
\address{Instituto de F\'{\i}sica Gleb Wataghin, UNICAMP,
         Campinas, SP 13083-970, Brazil}
\author{C. Salles}

\address{Dept. of Materials Science and Engineering, \\
         University of Wisconsin, Madison, WI 53706, USA}
\date{\today}
\maketitle

\begin{abstract}
We show that the simplest assumptions for the dynamics of particle production
allow us to understand the fluxes of hadrons and photons at mountain altitudes
as well as the structure of individual events. The analysis requires a heavy
nuclear component of primary cosmic rays above the ``knee" in the spectrum with
average mass number $<A> = 7.3 \pm 0.9$.
\end{abstract}

\section{Introduction}

The energy spectrum and chemical composition of primary cosmic rays have been
determined from direct observation above the earth's atmosphere using balloons
and spacecraft~\cite{shibata}. The technique is limited by the small size of
the detectors and the short exposure times. As a result of the steep energy
spectrum no direct observations are available above a primary energy of
roughly $10^{14}$~eV. In the interesting region of the ``knee" in the spectrum
and above information on composition has to be inferred from indirect
measurements of air showers at sea level or mountain altitudes~\cite{gaisser},
by using  large area detectors for long periods of time.

In this paper we infer the composition of the cosmic rays from measurements
at  mountain altitudes of the hadronic and electromagnetic component of the
air  cascades initiated at the top of the atmosphere.
A connection between the nature of the primary particle and air shower
observations requires the detailed understanding of particle interactions at
very high energies and forward scattering angles where no information is
available from accelerator-based experiments.
The basic problem is that one is faced with the  impossibility of deducing
two unknowns, the composition and the dynamics of  particle interactions,
from a single measurement. We nevertheless pursue this  challenge because
we are confident that we understand particle interactions  with sufficient
accuracy to meaningfully approach this problem.
We have indeed  formulated a model which is based on the most
straightforward assumptions and which respects the spirit of quantum
chromodynamics~\cite{costa}.
More importantly, it describes in quantitative detail single events, i.e.
shower
cores in their early stage of development, observed in emulsion chamber
experiments at mountain altitudes~\cite{lattes}. Here we will show that this
model describes the observed hadronic and electromagnetic spectrum at
mountain  altitudes, provided the mass number of the primary cosmic rays
is $<A> = 7.3 \pm  0.9$.
This is consistent with the result obtained by other indirect means.

Our model of particle production~\cite{costa} is guided by the features
of QCD-inspired models: approximate Feynman scaling in the
fragmentation region and an inelasticity slowly varying with energy.
The  rapidity density of secondary charged pions is parametrized as

\begin{equation}
{dN\over dy} = x {dN\over dx} = a {(1-x)^n \over{x}} \;,
\label{rapdens}
\end{equation}
where $y$ is the rapidity of the secondaries and the Feynman variable $x$ is
given by the ratio of the energy $E$ of the secondary particle to the incident
energy $E_0$. With $a = 0.12$ and $n = 2.6$ ($n = 3$ is expected on the
basis  of  counting rules) the overall features of the hadronic component
of single  events detected in emulsion chambers were quantitatively
reproduced. For  illustration, we present in Fig.~\ref{fig:families}
the hadronic integral spectrum of two events detected by the Brazil-Japan
Collaboration at  Mt. Chacaltaya, Bolivia (atmospheric depth
$540$ g/cm$^2$)~\cite{chinellato,bjc-1}, which are successfully
described by our  model~\cite{fluctuations}.
In the present paper we use the same model to calculate both the
hadronic and electromagnetic integral spectra of atmospheric
showers detected in large emulsion chamber experiments.
The calculation is performed by solving the cosmic-ray diffusion
equations using the rapidity distribution for particle production
given by Eq.~(\ref{rapdens}). Starting with the measured
all-particle primary spectrum at the top of the atmosphere, we propagate
the  particle showers down to the mountain altitude detection levels of
the various  experiments and investigate our results as a  function of
the assumed average  composition of the primary cosmic radiation.

\section{Hadronic and Electromagnetic Showers in the Atmosphere}

The flux of cosmic ray nucleons at the top of atmosphere (depth $t=0$) is
parametrized by a power-law spectrum

\begin{equation}
F_n (E,t=0) = N_0 \, E^{-(\gamma +1)} \;.
\label{primary-n}
\end{equation}
At this point no secondaries have been produced, hence the boundary condition
for the pionic component of the shower: $F_{\pi} (E,t=0) = 0$. The hadronic
flux $F_h (E,t)$ can be calculated for any depth $t=z$ in terms of the
interaction mean free path (MFP) of nucleons ($n$) and pions ($\pi$),
$\lambda_i
(E)$ with $i= n, \pi$. The result is~\cite{bellandi-1}

\begin{equation}
F_h (E,z) = F_n (E,z) + F_{\pi} (E,z) \;,
\label{flux-h}
\end{equation}
with
\begin{eqnarray}
F_n(E,z) = &&
 N_0 \, E^{-(\gamma +1)} \, e^{ -z H_n(E)} \;,
\label{flux-n} \\
F_{\pi}(E,z) = &&
 N_0 \, E^{-(\gamma +1)} {g_n(\gamma,E) \over H_n(E) -H_{\pi}(E)}
\nonumber \\ && \times \,
\left( e^{ -z H_{\pi}(E)} - e^{ -z H_n(E)} \right) \;.
\label{flux-p}
\end{eqnarray}
The dependence of the functions $H_i(E)$ and $g_i(\gamma,E)$ on the
rapidity  distribution, Eq.~(\ref{rapdens}), and on the MFP $\lambda_i(E)$,
is described in the Appendix.

The electromagnetic component of the shower is initiated by gamma rays from
the decay of the neutral pion, $\pi^{o} \rightarrow 2 \gamma$. With equal
multiplicity of $\pi^{+}$, $\pi^{-}$ and $\pi^{o}$ secondaries, the number of
neutral pions is half the number of the charged pions which is given by
Eq.~(\ref{flux-p}). The gamma-ray distribution is given by two-body
decay~\cite{navia}

\begin{equation}
{\cal F}_{\gamma} (E_{\gamma},z) dE_{\gamma} dz =
  2 \; \int_{E_{\gamma}}^{\infty} {dE \over E}
  \; \frac{1}{2} F_{\pi}(E,z) \, dE_{\gamma} dz \; .
\label{rate-g}
\end{equation}

A gamma ray with energy $E_{\gamma}$ at depth $z$ contributes to the
electromagnetic cascade $F_{\gamma}(E,t)$ with

\begin{eqnarray}
F_{\gamma}(E,t) = &&
 \int_{0}^{t} dz \int_{E_{\gamma}}^{\infty} dE_{\gamma}
 \; {\cal F}_{\gamma} (E_{\gamma},z) \nonumber \\
 && \times \, (e \, + \, \gamma)(E_{\gamma},E,t-z) \;.
\label{flux-g}
\end{eqnarray}
Here $(e \, + \, \gamma)(E_{\gamma},E,t-z)$ represents the photons
and  $e^+e^-$~pairs in the cascade produced by the photon with energy
$E_{\gamma}$.
We compute it in approximation A~\cite{nishimura} using the operator
formalism~\cite{bellandi-2}. The result is of the form of Eq.~(\ref{flux-g})
with $(e \, + \, \gamma)(E_{\gamma},E,t-z)$ given by the eigenvalues of
the electromagnetic cascade equations; see Appendix.

We are now ready to compute the integral energy spectrum for hadrons and
photons ($i=h, \gamma$),
\begin{equation}
I_i (>E,z) = \int_{E}^{\infty} F_i (E',z) \; dE' \; ,
\label{integral-i}
\end{equation}
which can be confronted with experimental results.

\section{Inelastic Cross Section for Hadron-Air Interactions}

Before proceeding with the analysis, it is necessary to describe the energy
dependence of the MFP which is inversely proportional to the inelastic cross
section for hadron-air interactions, i.e.

\begin{equation}
\lambda_i (E) = {24,100\rm (g/cm^2) \over \sigma_{in}^{i-air} (in\ mb)} \; .
\label{lambda-i}
\end{equation}
We calculated the inelastic cross section using the event generator
SIBYLL~\cite{sibyll}. The result can be parametrized by

\begin{equation}
{\rm\sigma_{in}^{i-air}} =
  s_i \, \left[1+b_i \ln^2 \left(\frac{E}{E_0}\right)\right] \; ,
\label{sigma-i}
\end{equation}
with $E_0=200$ GeV. Our results with $s_n=284.5$ mb and $b_n=3.852 \times
10^{-3}$ for proton-air and $s_{\pi}=211.0$ mb and $b_{\pi}=5.827 \times
10^{-3}$ for pion-air scattering are shown in Fig.~\ref{fig:sibyll}.

\section{Primary Composition and Shower Energy Spectra}

Having constructed an explicit model of particle interactions which
successfully describes individual events, see Fig.~\ref{fig:families}, we can
compute the flux of hadrons and photons at mountain altitude as a function of
the primary cosmic ray flux. For the primary spectrum we use a
parametrization~\cite{akeno} which is accurate in the energy region between
$300$ and $10^6$ TeV$/$particle relevant to our calculation. It accurately
extrapolates to lower energy measurements obtained with the Proton satellite
and the JACEE balloon flights~\cite{akeno,jacee-1}. The all-particle
differential energy spectrum, is given by

\begin{equation}
F_{all} (E,t=0) = (4.55 \pm 0.45) \times 10^{-11}
        \left[ {E \over 10^{3.67}} \right]^{-(\gamma +1)} \; ,
\label{primary-all}
\end{equation}
in units of $\rm(m^2\;sec\;sr\;TeV/particle)^{-1}$, with
$\gamma = 1.62 \pm 0.12$ below and
$\gamma = 2.02 \pm 0.05$ above $10^{3.67}$  TeV/particle.
Eq.~(\ref{primary-all})  describes the change in the slope of the spectrum
at the energy region known as  ``the knee''.
We parametrize our ignorance of the chemical composition of the primary
flux in terms of a single parameter $<A>$, the average mass number of
the primary nuclei. Heavy primaries are included in our formalism using
superposition~\cite{gaisser} in Eq.~(\ref{primary-n}). The projectile nucleus
of energy $E_0$ is considered to be the superposition of  $A$ nucleons
interacting independently, each having energy $E_0/A$. Although a
simplification, this assumption is quite acceptable, as long as the nuclei
fragment relatively rapidly.

We first calculate the integral energy spectra of electromagnetic showers,
Eq.~(\ref{integral-i}), at the detection level of Mt. Chacaltaya, using the
extremes values for $<A>$ corresponding to pure proton ($A=1$) and to
pure iron  ($A=56$). The results are shown in Fig.~\ref{fig:chacaltaya},
by the dashed curves. The predictions bracket the experimental
data~\cite{bjc-2}.
That a pure proton spectrum cannot describe these results is not totally
surprizing~\cite{shibata-ren}.
It is well known that the relatively low rate of detected gamma-ray
families (and also of halo families) cannot be understood
in  models with approximate Feynman scaling unless heavy primaries
contribute to  the cosmic ray flux. We subsequently determine, by
chi-square minimization,  the average mass number that best describes
the data. We obtain $<A> = 7.3 \pm 0.9$ (solid line in
Fig.~\ref{fig:chacaltaya}).

Having fixed all parameters, we can confront the model with any other
observations. We find that it successfully describes both the hadronic
(Fig.~\ref{fig:fujikanb}a) and electromagnetic
(Fig.~\ref{fig:fujikanb}b) components of the atmospheric showers detected in
large emulsion chambers at Mt. Fuji~\cite{fuji} in Japan (atmospheric depth
650  g/cm$^2$), and at Mt. Kanbala~\cite{kanbala} in China (520 g/cm$^2$).

\section{Conclusions}

We conclude that with the simplest assumptions for the production of
secondaries based on approximate scaling in the fragmentation region, it is
possible to explain a broad set of experimental data on very high energy
cosmic rays in the atmosphere, namely the lateral spread and the integral
spectra of superfamilies (as in Ref.~\cite{costa} and
Fig.~\ref{fig:families}),
and the energy spectra of hadronic and electromagnetic showers detected
in large emulsion chamber experiments (as in Figs.~\ref{fig:chacaltaya}
and~\ref{fig:fujikanb}).
This scenario requires a  primary composition with average mass number
$7.3 \pm 0.9$. We investigated the sensitivity of this quantity to different
parametrizations of the all-particle  spectrum ~\cite{jacee-2,biermann} and
the best fit invariably yields $<A> \simeq  7$.
Our  result is also consistent with underground muon
measurements~\cite{macro} which yield an $A$-value of
$ 10 \pm 4$  in the 1 to 1,000 TeV energy range.

It has been noted elsewhere~\cite{yuda-halzen} that it is difficult to
establish
whether one must adopt a heavy primary composition along with a model of
particle production based on scaling or, alternatively, a proton dominant
composition along with a strong violation of Feynman scaling. It should be
noted however that in our analysis the particle interaction model was
determined on the basis of an independent study of single events initiated by
protons deep in the atmosphere. The a posteriori analysis of the hadron and
photon spectra at mountain altitude presented here, required the introduction
of heavy primaries.

\acknowledgements

This research was supported in part by the Conselho Nacional de
Desenvolvimento Cient\'{\i}fico e Tecnol\'{o}gico (CNPq, Brazil), in
part by the U.S.~Department of Energy under Grant No.~DE-FG02-95ER40896,
and by the National Science Foundation under Contract No.~DMR-9319421,
and in part by the University of Wisconsin Research Committee
with funds granted by the Wisconsin Alumni Research Foundation.

\appendix
\section*{Definitions in the Analytical Solutions}
Complete definition of the hadronic flux components presented
in  Eqs.~(\ref{flux-n}) and~(\ref{flux-p})
requires the following functions (for $i=n,\pi$):

\begin{eqnarray}
H_n(E) = && \frac{1}{\lambda_n(E)} \,
 \left(1-\frac{1}{2} \left< \sigma_n^{\gamma}\right> \right)
 - \frac{1}{2} \left< {\sigma_n^{\gamma} \over
   \lambda_n(E/\sigma_n)}\right> \;, \\
H_{\pi}(E) = && \frac{1}{\lambda_{\pi}(E)} \,
 \left(1+\frac{1}{2} \left< \sigma_{\pi}^{\gamma} \right>
 + \frac{1}{2} g_{\pi}(\gamma) \right) \nonumber \\ &&
 - \, \frac{3}{2} \left< {\sigma_{\pi}^{\gamma} \over
 \lambda_{\pi}(E/\sigma_{\pi})} \right>
 - g_n(\gamma,E) \;, \\
g_i(\gamma,E) = &&
 \int_0^1 \frac{1}{\lambda_i(E/x)} {dN \over dx} \, x^{\gamma} \, dx \;,
\end{eqnarray}
with

\begin{eqnarray*}
g_i(\gamma) = && \int_0^1 {dN \over dx} \, x^{\gamma} \, dx \;, \\
\left< \sigma_i^{\gamma}\right> =
 && \int_0^1 f(\sigma_i) \sigma_i^{\gamma} d\sigma_i \;, \\
\left< {\sigma_i^{\gamma} \over \lambda_i(E/\sigma_i)}\right> =
 && \int_0^1 \frac{1}{\lambda_i(E/\sigma_i)}
 f(\sigma_i) \sigma_i^{\gamma} d\sigma_i \; .
\end{eqnarray*}

The elasticity distribution is assumed to be

\begin{equation}
f(\sigma_i) = \left(1+\beta \right) \, \left(1-\sigma_i \right)^{\beta} \; ,
\label{elasticity}
\end{equation}
where $\beta$ fulfills a consistency relation between average elasticity
$\left<\sigma\right>$ and average inelasticity $\left< K \right>$, so that
$\left<\sigma\right> \, + \, \left< K \right> \, = \, 1$ (energy conservation).
The eigenvalues for the electromagnetic cascade equations are~\cite{bellandi-2}

\begin{eqnarray}
\Pi_{\gamma}(s,t) = && \frac{1}{X_0}
{B(s) \over (\lambda_1(s) - \lambda_2(s))}
\nonumber \\ && \times \,
\left( e^{ \lambda_1(s) \, t /{X_0}}
    -  e^{ \lambda_2(s) \, t /{X_0}} \right) \;, \\
\Gamma_{\gamma}(s,t) = && \frac{1}{X_0}
    \left( H_2(s) \; e^{ \lambda_1(s) \, t /{X_0}}\right.
\nonumber \\ && + \,
   \left. H_1(s) \; e^{ \lambda_2(s) \, t /{X_0}} \right) \;,
\end{eqnarray}
where $H_1(s), H_2(s), \lambda_1(s), \lambda_2(s)$ and $X_0$ are parameters
with standardized definitions in cascade theory~\cite{nishimura}.
Subsequently  $(e \, + \, \gamma)(E_{\gamma},E,t-z)$
in Eq.~(\ref{flux-g}) should be replaced by
$\left( \Pi_{\gamma} (s,t-z) + \Gamma_{\gamma} (s,t-z) \right)$
with $s$ evaluated at the pole $s=\gamma$.

\newpage
\begin{figure}
\caption{Integral energy spectra of hadronic superfamily events
detected at Mt.~Chacaltaya~[5,6].
Ursa Maior event ($\diamondsuit$) and Centauro\,VII data
($\bigtriangleup$), are compared to the calculation
of Ref.~[3] (solid line), using the $x$-distribution of Eq.~(1).
For illustrative purposes, the data of Centauro\,VII have been
shifted by a factor 10.}
\label{fig:families}
\end{figure}

\begin{figure}
\caption{Inelastic cross sections for $p-$air ($\bigcirc$) and
$\pi-$air ($\Box$) scattering computed using SIBYLL~[12]
and parameterized by Eq.~(10).}
\label{fig:sibyll}
\end{figure}

\begin{figure}
\caption{Integral energy spectra of electromagnetic showers
($\diamondsuit$),detected at Mt.~Chacaltaya~[15], compared
to the calculation using the $x$-distribution of Eq.~(1).
Dashed lines: $<A>= 1$ (proton) and $56$ (iron);
solid line: $<A>=7.3 \pm 0.9$, dotted lines: calculated
from uncertainties in $A$ and $\gamma$.}
\label{fig:chacaltaya}
\end{figure}

\begin{figure}
\caption{Integral energy spectra of showers detected at
Mt.~Fuji~[17] ($\bigcirc$) and Mt. Kanbala~[18]
($\bigtriangleup$), compared to the analytical calculation
using the $x$-distribution of Eq.~(1), with
$<A>= 7.3 \pm 0.9$ (solid line). (a) Hadron induced showers;
(b) Electromagnetic showers.
The data of Mt. Kanbala have been shifted by a factor 100.
Dotted lines are calculated from
uncertainties in $A$ and $\gamma$.}
\label{fig:fujikanb}
\end{figure}


\newpage
\pagestyle{empty}

\vspace*{8.5in}
\includegraphics{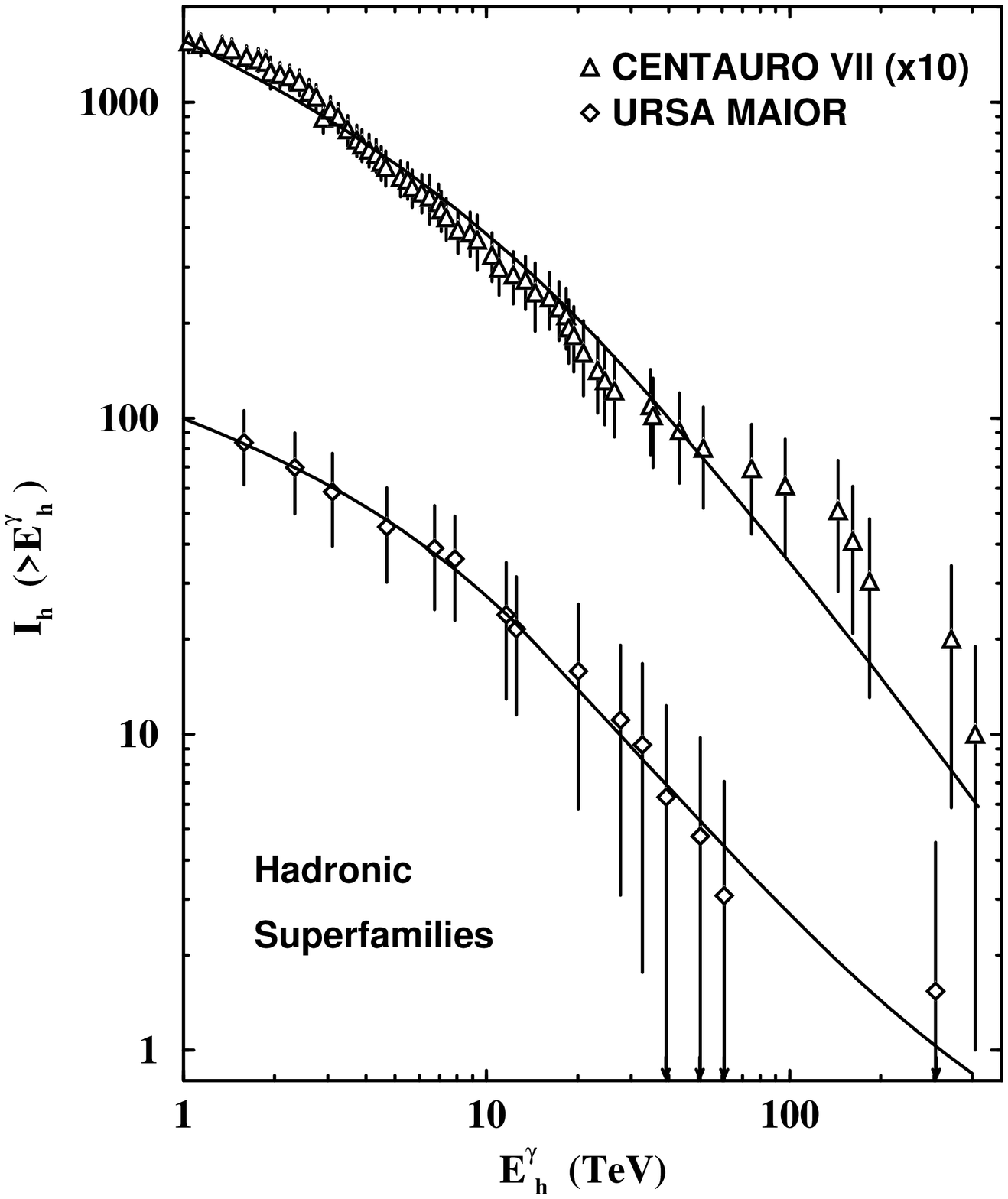}
\centerline{\quad\qquad\bf Fig.~1}

\newpage
\vspace*{8.5in}
\includegraphics{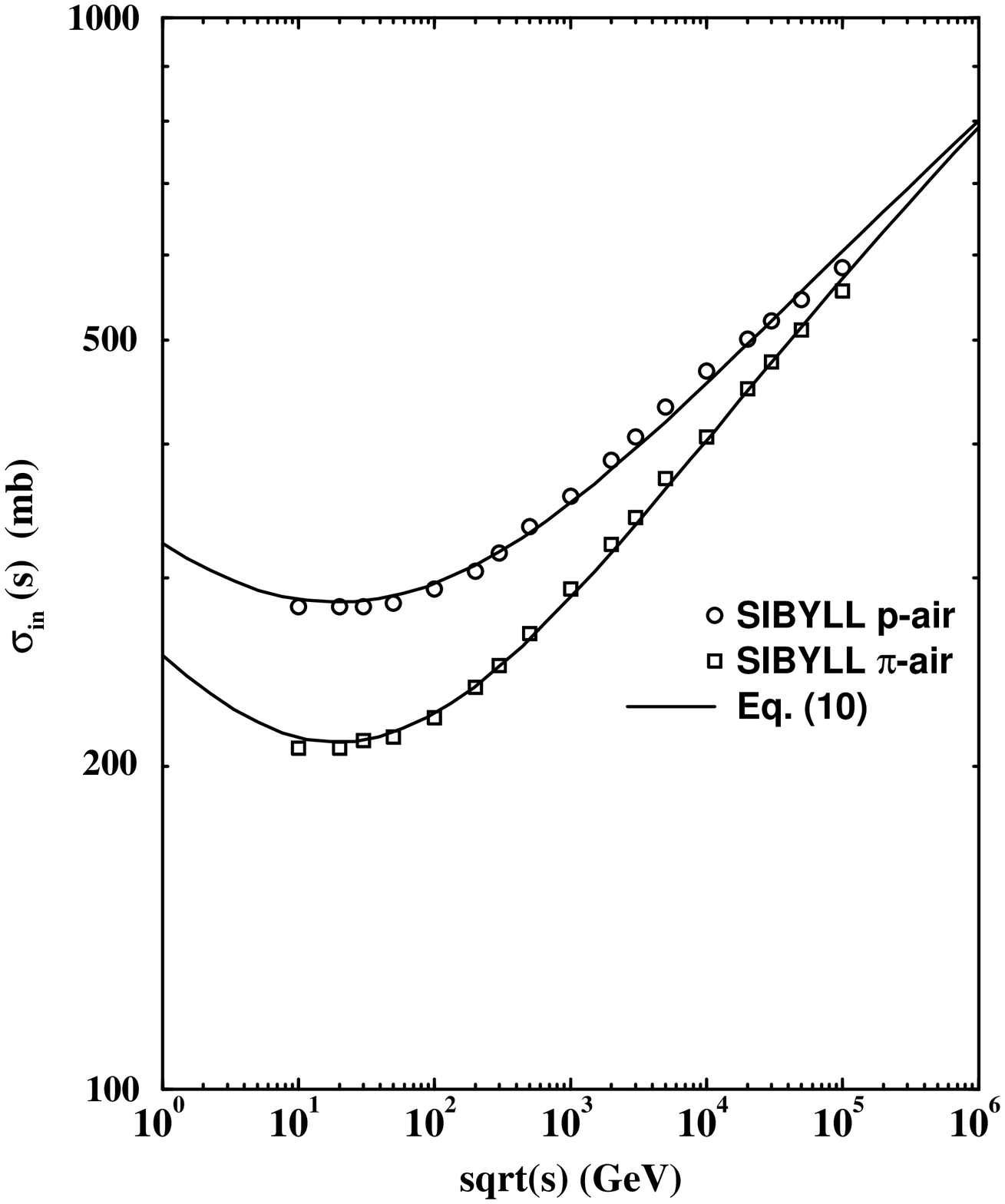}
\centerline{\quad\qquad\bf Fig.~2}

\newpage
\vspace*{8.5in}
\includegraphics{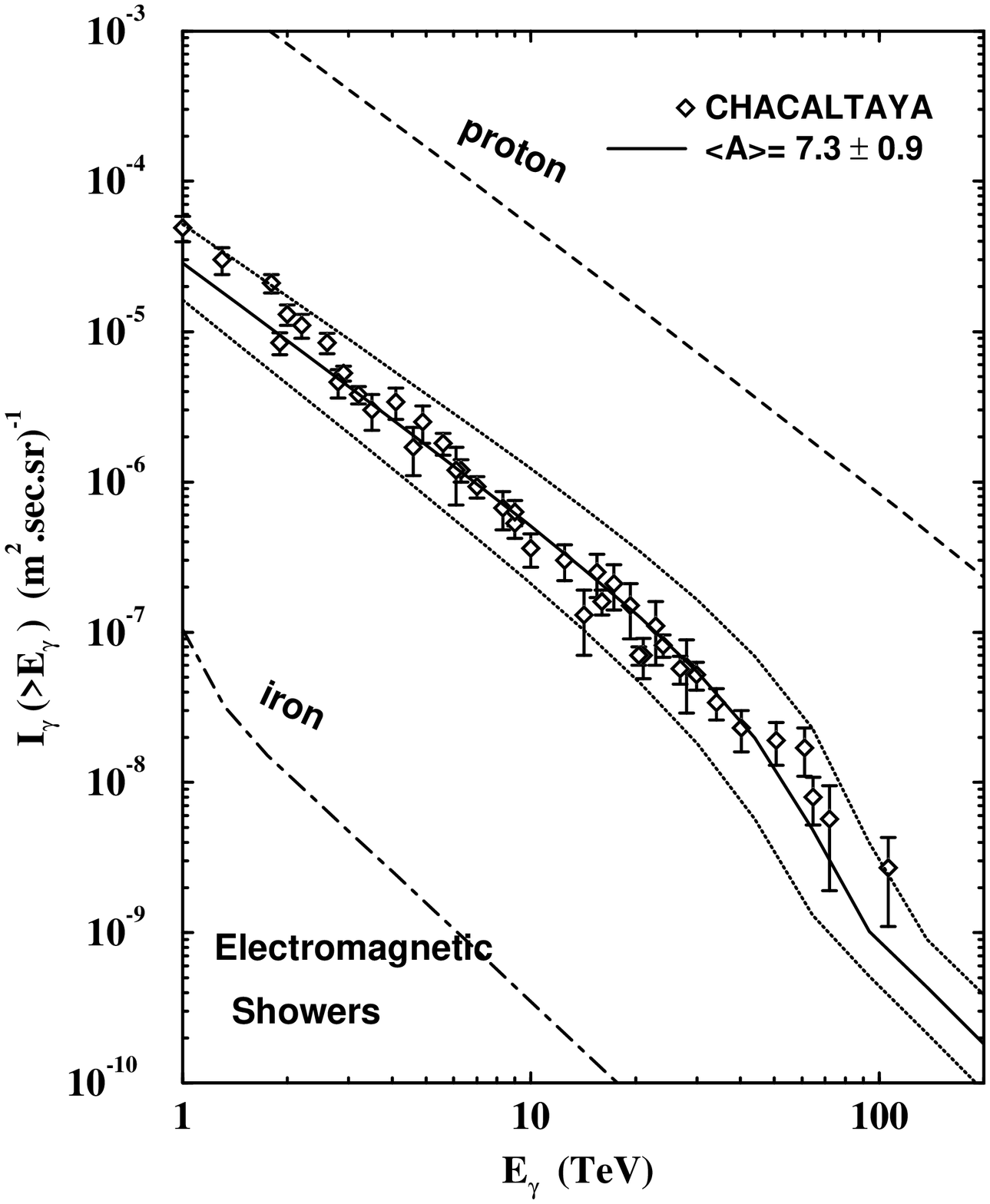}
\centerline{\quad\qquad\bf Fig.~3}


\newpage
\vspace*{8.5in}
\includegraphics{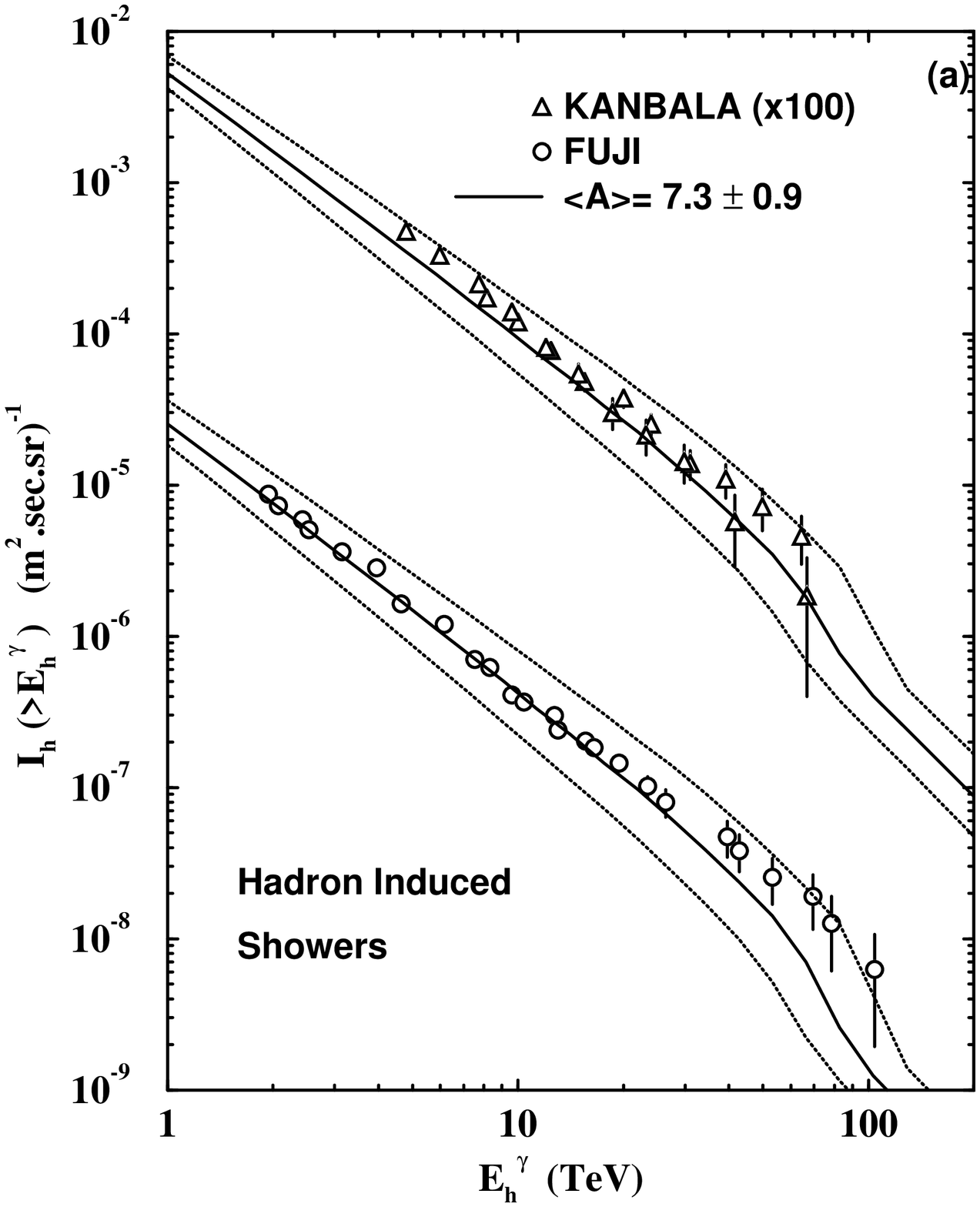}
\centerline{\quad\qquad\bf Fig.~4a}

\newpage
\vspace*{8.5in}
\includegraphics{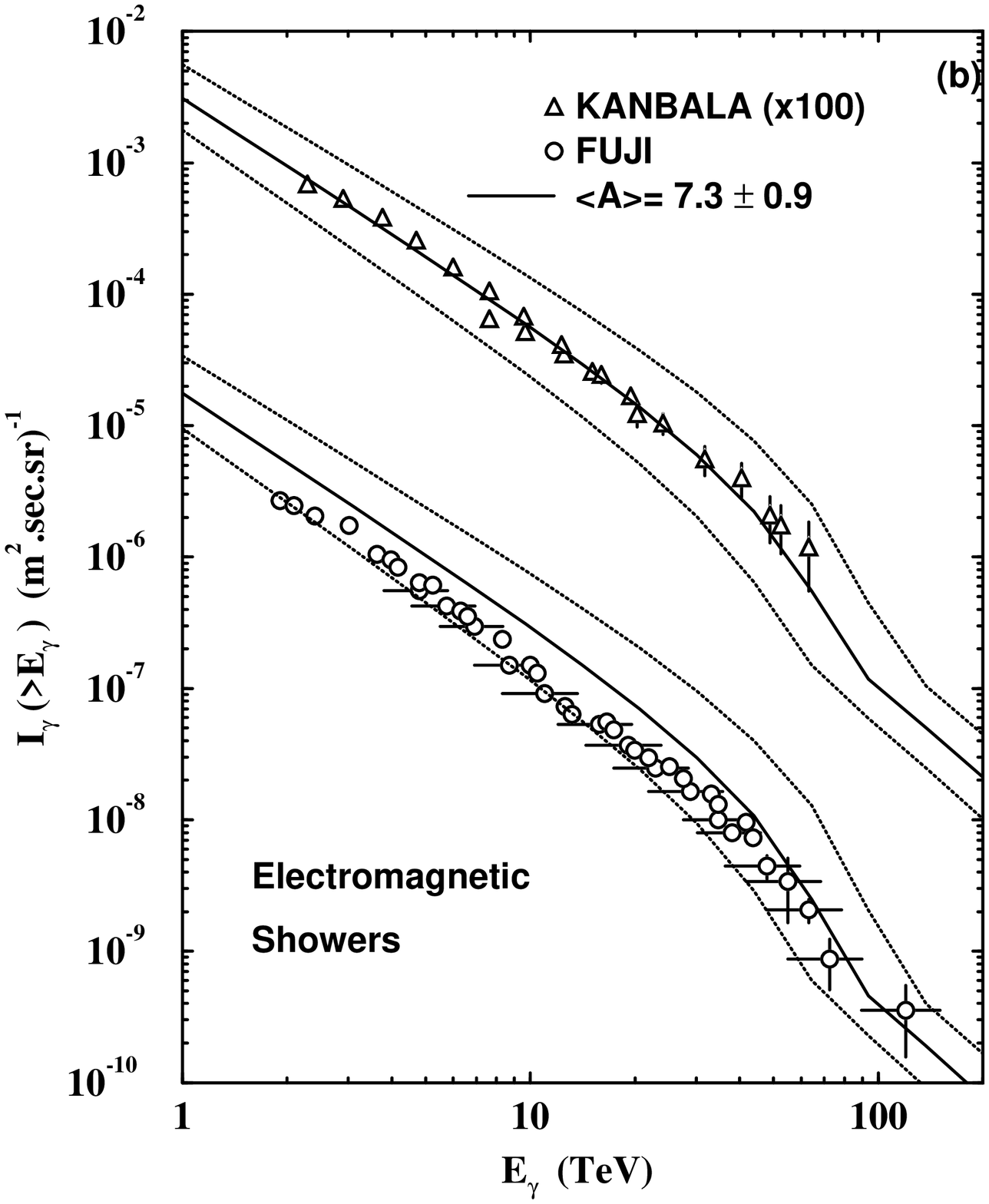}
\centerline{\quad\qquad\bf Fig.~4b}


\begin{references}
\addtolength{\itemsep}{-0.05in}
\frenchspacing

\bibitem{shibata}
For a recent review, see T. Shibata, in
{\it Rapporteur Talk presented at the 24th International
Cosmic Ray Conference}
(Rome, Italy, 1995), University of Tokyo ICRR-Report-343-95-9, 1995.

\bibitem{gaisser}
T.K. Gaisser, in {\it Cosmic Rays and Particle Physics},
(Cambridge University Press, Cambridge, 1992).

\bibitem{costa}
C.G.S. Costa, F. Halzen and C. Salles,
Phys. Rev. D {\bf 52}, 3890 (1995).

\bibitem{lattes}
C.M.G. Lattes, Y. Fujimoto and S. Hasegawa,
Phys. Rep. {\bf 65}, 151 (1980).

\bibitem{chinellato}
J.A. Chinellato, Ph.D Thesis, Universidade Estadual de Campinas,
Brazil (1981).

\bibitem{bjc-1}
Brazil-Japan Collaboration, J.A.~Chinellato {\it et al.}, in {\it Proc.
of the 21st International Cosmic Ray Conference}, Adelaide, Australia,
ed. by R.J.~Protheroe (Graphic Services, Northfield, 1990),
Vol.~8, p.~259.

\bibitem{fluctuations}
The drawback of using analytic shower theory is the neglect of fluctuations.
This is not a problem here as their effect is masked by the fact that
average  depth and energy of the incident particle are fitted as parameters.
These should have acceptable values~\cite{costa}, but cannot be rigorously
interpreted on an event by event basis. The procedure is equivalent to
averaging several events.

\bibitem{bellandi-1}
J. Bellandi {\it et al.},
Nuovo Cimento C {\bf 14}, 15 (1991).

\bibitem{navia}
C.E. Navia {\it et al.},
Phys. Rev. D {\bf 40}, 2898 (1989).

\bibitem{nishimura}
J. Nishimura, in {\it Handbuch der Physik}, ed. by K. Sitte
(Springer, Berlin, 1967), Vol.~62, Part~2, p.~1 .

\bibitem{bellandi-2}
J. Bellandi Fo. {\it et al.},
J. Phys. A: Math. Gen. {\bf 25}, 877 (1992).

\bibitem{sibyll}
R.S. Fletcher, T.K. Gaisser, P. Lipari and T. Stanev,
Phys. Rev. D {\bf 50}, 5710 (1994).

\bibitem{akeno}
T. Hara {\it et al.}, in
{\it Proc. of the 18th International Cosmic Ray Conference},
Bangalore, India, 1983, ed. by N. Durgaprasad {\it et al.}
(CTIFR, Bombay, 1983), Vol.~9, p.~198;
see also, M. Nagano {\it et al.},  J. Phys. G {\bf 10}, 1295 (1984).

\bibitem{jacee-1}
JACEE Collaboration, K. Asakimori {\it et al.}, in
{\it Proc. of the 7th International Symposium on Very High Energy
Cosmic Ray  Interactions}, Tokyo, Japan, ed. by Y.~Fujimoto {\it et al},
(Tokyo, 1994), p.~513.

\bibitem{bjc-2}
Brazil-Japan Collaboration, C.M.G. Lattes {\it et al.},
Prog. Theor. Phys. Suppl. {\bf 47}, 1 (1971);
and in
{\it Proc. of the 13th International Cosmic Ray Conference},
Denver, USA, (Colorado Associated Univ. Press, Boulder, 1973),
Vol.~3, p.~2219.

\bibitem{shibata-ren}
M. Shibata, Phys. Rev. D {\bf 24}, 1847 (1981);
J.R. Ren {\it et al.}, Phys. Rev. D {\bf 38}, 1404 (1988).

\bibitem{fuji}
Mt. Fuji Collaboration, M. Akashi {\it et al.},
Nuovo Cimento A {\bf 65}, 355 (1981).

\bibitem{kanbala}
China-Japan Emulsion Chamber Collaboration, J.R. Ren {\it et al.},
Nuovo Cimento C {\bf 10}, 43 (1987).

\bibitem{jacee-2}
JACEE Collaboration, T.H. Burnett {\it et al.}, in
{\it Proc. of the International Symposium on Cosmic Ray
and Particle Physics}, Tokyo, Japan, ed. by A.~Ohsawa and T.~Yuda
(Yamada, Tokyo, 1984), p.~468.

\bibitem{biermann}
B. Wiebel-Sooth, P.L. Biermann and H. Meyer, in
{\it Proc. of the 24th International Cosmic Ray Conference},
Rome, Italy (INFN, Rome, 1995), Vol.~2, p.~656.

\bibitem{macro}
MACRO Collaboration, in
{\it Proc. of the 24th International Cosmic Ray Conference},
Rome, Italy (INFN, Rome, 1995),  Vol.~2, p.~689.

\bibitem{yuda-halzen}
T. Yuda, in
{\it Rapporteur Talk presented at the 22nd International
Cosmic Ray Conference},
(Dublin, Ireland, 1991), Vol.~5, p.~313;
F. Halzen, in {\it Proc. of  VII International Symposium on Very High
Eenergy Cosmic Ray Interactions}, Ann Arbor, USA, ed. by L. Jones
(AIP, New York, 1993), AIP Conf. Proc. Vol.~276, p.~679.

\end{references}
\end{document}